\documentclass[%
 reprint,
superscriptaddress,
amsmath,amssymb,
aps]{revtex4-2}

\usepackage{graphicx}
\usepackage{float}
\usepackage{dcolumn}
\usepackage{bm}
\usepackage{physics}
\usepackage{xcolor}
\usepackage{siunitx}
\usepackage{multirow}
\usepackage{ulem}
\usepackage{comment}

\date{November 2025}
\definecolor{bl}{rgb}{0, .1, .6}
\usepackage[colorlinks=true, citecolor = bl, linkcolor = bl, urlcolor=bl, pdfborder={0 0 0}]{hyperref}

\newcommand{\gf}[1]{{\color{orange} \textbf{GF: } #1}}
\newcommand{\pel}[1]{{\color{cyan} \textbf{PL: } #1}}
\newcommand{\af}[1]{{\color{blue} \textbf{AF: } #1}}
\newcommand{\fjr}[1]{{\color{green} \textbf{FR: } #1}}

\DeclareSIUnit\gauss{G}

\usepackage{xcolor}

\begin{document}
	\title{High-resolution spectroscopy of $^{162}$Dy Rydberg levels}    

    \author{G. Ferioli}
	\affiliation{Dipartimento di Fisica e Astronomia, Universit\`a degli Studi di Firenze, Via G. Sansone 1, 50019 Sesto Fiorentino, Italy}   
    \affiliation{European Laboratory for Nonlinear Spectroscopy (LENS), Via N. Carrara 1, 50019 Sesto Fiorentino, Italy}

    \author{P. Lombardi}
    \affiliation{Istituto Nazionale di Ottica, Consiglio Nazionale delle Ricerche (CNR-INO), Via N. Carrara 1, 50019 Sesto Fiorentino, Italy}
    \affiliation{European Laboratory for Nonlinear Spectroscopy (LENS), Via N. Carrara 1, 50019 Sesto Fiorentino, Italy}

    \author{P. Sekhar}
	\affiliation{Dipartimento di Fisica e Astronomia, Universit\`a degli Studi di Firenze, Via G. Sansone 1, 50019 Sesto Fiorentino, Italy}   
    \affiliation{European Laboratory for Nonlinear Spectroscopy (LENS), Via N. Carrara 1, 50019 Sesto Fiorentino, Italy}
   
    \author{E. Sol{\'e} Cardona}
    \affiliation{Dipartimento di Fisica e Astronomia, Universit\`a degli Studi di Firenze, Via G. Sansone 1, 50019 Sesto Fiorentino, Italy}   
    \affiliation{Istituto Nazionale di Ottica, Consiglio Nazionale delle Ricerche (CNR-INO), Via Moruzzi 1, 56124 Pisa, Italy}
    
    \author{N. Preti}
    \affiliation{Dipartimento di Fisica e Astronomia, Universit\`a degli Studi di Firenze, Via G. Sansone 1, 50019 Sesto Fiorentino, Italy}  
    \affiliation{European Laboratory for Nonlinear Spectroscopy (LENS), Via N. Carrara 1, 50019 Sesto Fiorentino, Italy}
    \affiliation{Istituto Nazionale di Ottica, Consiglio Nazionale delle Ricerche (CNR-INO), Via Moruzzi 1, 56124 Pisa, Italy}
    
    \author{C. Drevon}
    \affiliation{Dipartimento di Fisica e Astronomia, Universit\`a degli Studi di Firenze, Via G. Sansone 1, 50019 Sesto Fiorentino, Italy}  
    \affiliation{European Laboratory for Nonlinear Spectroscopy (LENS), Via N. Carrara 1, 50019 Sesto Fiorentino, Italy}
    \affiliation{Istituto Nazionale di Ottica, Consiglio Nazionale delle Ricerche (CNR-INO), Via Moruzzi 1, 56124 Pisa, Italy}

    \author{N. Antolini}
    \affiliation{Istituto Nazionale di Ottica, Consiglio Nazionale delle Ricerche (CNR-INO), Via Moruzzi 1, 56124 Pisa, Italy}
    \affiliation{European Laboratory for Nonlinear Spectroscopy (LENS), Via N. Carrara 1, 50019 Sesto Fiorentino, Italy}

     \author{L. Tanzi}
    \affiliation{Istituto Nazionale di Ottica, Consiglio Nazionale delle Ricerche (CNR-INO), Via N. Carrara 1, 50019 Sesto Fiorentino, Italy}
    \affiliation{European Laboratory for Nonlinear Spectroscopy (LENS), Via N. Carrara 1, 50019 Sesto Fiorentino, Italy}
    
    \author{G. Modugno}
    \affiliation{Dipartimento di Fisica e Astronomia, Universit\`a degli Studi di Firenze, Via G. Sansone 1, 50019 Sesto Fiorentino, Italy}   
    \affiliation{European Laboratory for Nonlinear Spectroscopy (LENS), Via N. Carrara 1, 50019 Sesto Fiorentino, Italy}
   \affiliation{Istituto Nazionale di Ottica, Consiglio Nazionale delle Ricerche (CNR-INO), Via Moruzzi 1, 56124 Pisa, Italy}

    \author{C. Gabbanini}
    \affiliation{Istituto Nazionale di Ottica, Consiglio Nazionale delle Ricerche (CNR-INO), Via Moruzzi 1, 56124 Pisa, Italy}

    \author{F. Robicheaux}
	\affiliation{Department of Physics and Astronomy, Purdue University, West Lafayette, IN 47907, USA}
	\affiliation{Purdue Quantum Science and Engineering Institute, Purdue University, West Lafayette, IN 47907, USA}

    \author{A. Fioretti}
    \affiliation{Istituto Nazionale di Ottica, Consiglio Nazionale delle Ricerche (CNR-INO), Via Moruzzi 1, 56124 Pisa, Italy}
    \affiliation{European Laboratory for Nonlinear Spectroscopy (LENS), Via N. Carrara 1, 50019 Sesto Fiorentino, Italy}

\date{\today}
\begin{abstract}
Highly excited Rydberg states of lanthanides are a promising, yet largely unexplored, playground for quantum studies. Here, we report on the first high-resolution spectroscopy of  $^{162}$Dy obtained by two-color trap depletion spectroscopy in a magneto-optical trap. The absolute excitation frequency of over 700 states with effective principal quantum number $21\leq n\leq 130$ is measured with an  accuracy of 20 MHz. Most states are assigned to the 8  different  
series converging to the first $4f^{10}(^5I_8)6s (^2S_{1/2})\, J=17/2$ ionization potential. This energy is measured at $E_{\rm IP}=47901.8265\pm 0.0008\,$cm$^{-1}$,
 improving the precision of the literature value by over an order of magnitude. A multichannel quantum defect theory approach is used to benchmark and refine the assignments and to characterize six observed perturbing states belonging to higher ionization limits.
These results pave the way for using dysprosium in Rydberg-based quantum architectures, leveraging the unique properties arising from its complex electronic structure. They also represent a compelling benchmark for ab-initio calculations of open-shell atomic systems.

\end{abstract}

\maketitle

\section{Introduction}
\label{sec:intro}

Rydberg atoms are among the most promising platforms for quantum technologies~\cite{Adams_2020}, with wide-ranging applications in quantum information processing~\cite{saffman2010quantum}, quantum simulation~\cite{browaeys2020many}, quantum metrology \cite{dnorcia2019seconds, clock2019madjarov}, and quantum optics~\cite{Firstenberg_2016,Kumlin_2023}. 

While most experimental progress to date has relied on alkali atoms, the richer electronic structure of alkaline-earth and alkaline-earth-like species, such as strontium (Sr) and ytterbium (Yb), provides additional tools for cooling, trapping, and manipulating techniques \cite{alkaline2018cooper, narrow2019saskin, times2019covey, ytterbium2022jenkins, wilson2022trapping}. These features have enabled new approaches, including erasure conversion protocols~\cite{wu2022erasure}, mid-circuit operations~\cite{midcircuit2023lis, ma2023high}, and high-fidelity entangling gates~\cite{peper2025spectroscopy,endres2025highfidelity,senoo2025highfidelityentanglementcoherentmultiqubit}.

Lanthanide atoms with an open submerged 4f-shell, such as dysprosium (Dy) and erbium (Er), share many of these favorable properties, while also offering 
further/enhanced
possibilities for quantum control. Similarly to Sr and Yb systems, the presence of multiple closed optical transitions with widely varying linewidths (ranging from megahertz-broad to ultra-narrow, clock-like transitions) has enabled single-atom trapping~\cite{bloch2023trapping,grun2024optical}, high-fidelity imaging~\cite{bloch2023trapping,grun2025light,su2025fast}, cooling to the motional ground state~\cite{biagioni2025narrow}, and precise internal-state manipulation~\cite{patscheider2021observation,petersen2020spectroscopy}.

At the same time, the large electronic angular momentum of the ground state gives rise to a manifold of long-lived Zeeman sublevels that can serve as a resource for high-dimensional qudit encoding~\cite{kiktenko2020scalable,gonzalez2022hardware}, as originally proposed for holmium (Ho)~\cite{saffman2008scaling}, or for the generation of nonclassical spin-cat states~\cite{Kruckenhauser2025dark}. In addition, the large vector and tensor polarizabilities of the ground state enable the engineering of state-dependent trapping potentials over a broad range of wavelengths~\cite{du2024atomic}. This capability allows controlled coupling between internal and motional degrees of freedom and opens the door to improved motional gate schemes~\cite{erasure2025shaw}, with applications in the quantum simulation of fermionic systems~\cite{gozalez2023fermionic}.

Unlike Er~\cite{trautmann2021spectroscopy}, for which high-resolution Rydberg spectroscopy has recently become available, spectroscopic information on Rydberg states of Dy remains relatively limited. Earlier investigations were primarily based on multiphoton resonance ionization spectroscopy~\cite{Studer2017dysprosium}, which did not provide the level of precision now required for Rydberg-based quantum science applications. Here, we present the first high-resolution spectroscopic survey of Rydberg states in $^{162}$Dy, detecting more than 700 spectral lines with a typical accuracy of \SI{20}{\mega\hertz}. We further determine the ionization potential with an order-of-magnitude improvement in precision over earlier measurements. By analyzing the observed perturbations, we identify the character of the main Rydberg series and benchmark our assignments using multichannel quantum defect theory (MQDT).
\begin{figure*}[htbp]
\centering
\includegraphics[width=1\linewidth]{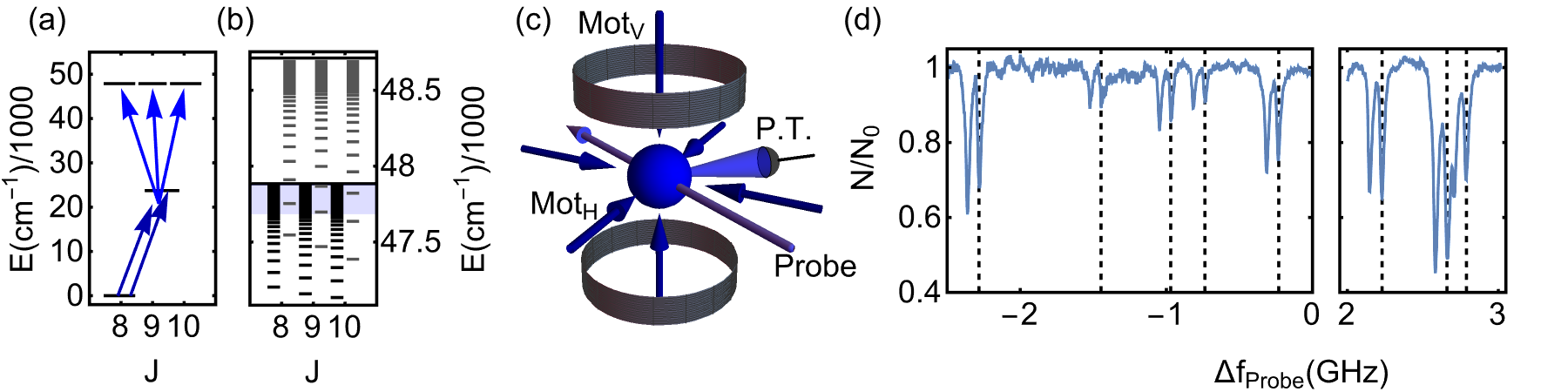}
\caption{(a) Two-photon excitation scheme to the Rydberg levels. The first photon, nearly resonant with the intermediate 
\(4f^{10}(^5I_8)6s6p(^1P^\circ_1)\) \(J=9\) level, corresponds to the transition used for MOT operation. The two arrows indicate the two different excitation pathways employed (see main text). Owing to selection rules, Rydberg states with \(J = 8, 9,\) and \(10\) can be accessed, as highlighted by the three arrows originating from the intermediate state.
(b) Relevant Rydberg series: black lines denote Rydberg levels converging to the first ionization threshold, while gray lines correspond to those associated with the second ionization threshold. The blue-shaded area indicates the range of frequencies we explored.
(c) Sketch of the core of the experimental platform. The two lasers used to operate the MOT are labeled $\mathrm{MOT}_V$ and $\mathrm{MOT}_H$, respectively. The laser exciting the atoms to the Rydberg states is labeled as Probe, and the MOT fluorescence is detected by a phototube (PT).
(d) Typical spectroscopic signal for \(n=90\). Over a frequency span of approximately \SI{6}{\giga\hertz}, eight distinct features corresponding to the expected transitions are clearly visible and marked by vertical dashed lines. Each feature exhibits a doublet structure arising from the two different MOT laser frequencies (see main text). {$\Delta f_{\text{Probe}}= 0$ corresponds to \SI{1435.657500(20)}{\tera\hertz}.}
}

\label{fig:levelscheme}
\end{figure*}

This article is organized as follows. In Sec.~\ref{sec:setup}, we describe the experimental setup and spectroscopy methods. In Sec.~\ref{sec:results}, we present the spectroscopic results, including the determination of the ionization potential $E_{\mathrm{IP}}$ and the assignment of the most relevant Rydberg series. In Sec.~\ref{sec:MQDT}, we outline the MQDT model used to fit the data. In Sec.~\ref{sec:amplitude}, we discuss the information extracted from the amplitudes of the depletion signals. Finally, Sec.~\ref{sec:conclusions} summarizes our conclusions and outlook.

\section{Experimental setup and methods}
\label{sec:setup}
{Dy Rydberg-state spectroscopy is performed in a magneto-optical trap (MOT) operating on the broad transition ($\Gamma = 2\pi \times \SI{32}{\mega\hertz}$, wavelength \SI{421}{\nano\meter}) connecting the ground state to the ${\rm 4f^{10}(^5I_8)6s6p(^1P^\circ_1)}\,J=9$ excited state. 

For technical reasons~\footnote{The 3D magneto-optical trap is loaded by a slow atomic beam outsourced from a 2D MOT that provides also the vertical beams of the 3D MOT, while the horizontal beams are provided by an independent laser. Optimal operation of the system is thus obtained for slightly different frequencies of the two lasers.}  and to maximize the number of trapped atoms, the MOT operates with two different detunings $\Delta$: one for the vertical beams ($\Delta \simeq -3\Gamma$) and another for the horizontal beams ($\Delta \simeq -\Gamma$). Under these conditions, the trap contains approximately $10^4$ atoms at a temperature of \SI{60}{\milli\kelvin}. The MOT lifetime is about \SI{5}{\milli\second}, limited by population trapping in dark states~\cite{Youn2010DyMOT421nm}. 

Excitation to Rydberg states is achieved via a two-photon transition, as illustrated in Fig.~\ref{fig:levelscheme}(a). The first photon is directly provided by the MOT light, while a second probe laser~\footnote{Toptica, mod. DLC TA-SHG PRO, softly focused to \SI{1}{\milli\meter} waist and nearly \SI{280}{\milli\watt} power on the atoms.} promotes atoms from the MOT excited state to highly-excited Rydberg states.  Under resonant conditions, this process introduces an additional loss channel for the MOT as a fraction of the trapped atoms is removed from the cooling cycle. As a result, Rydberg resonances manifest as a reduction of the MOT fluorescence, monitored with a photomultiplier tube while the probe laser is scanned. A similar technique has been used in the context of spectroscopy of Rydberg levels in Ho~\cite{Hostetter2015holmium} and Sr~\cite{couturier2019measurement}. {In Fig.\,\ref{fig:levelscheme}(b) we  sketch the frequency range we explored,  with a schematic representation of the Rydberg levels we expect to excite with our scheme, and in  Fig.\,\ref{fig:levelscheme}(c)} the schematic of the magneto-optical trap.

The energy of each Rydberg state is determined measuring the wavelength of the two lasers with a \SI{2}{\mega\hertz}-precision wavemeter~\footnote{High Finesse WS8-2} whose absolute accuracy is kept on the same order by periodic calibration against a Sr atomic reference. This protocol guaranties an absolute uncertainty of \SI{20}{\mega \hertz} over the whole explored energy range~\footnote{The absolute uncertainty characterizing the determination of the Rydberg levels is given by the wavemeter precision, the uncertainty with which we can locate the depletion peak, and the possible drift of the MOT laser during the spectroscopy campaign. The first and second terms are stochastic and independent errors, and combine into an overall contribution of around \SI{10}{\mega\hertz}. The latter term is again of the order of \SI{10}{\mega\hertz}, but because of its systematic origin, characterized by a rectangular distribution, needs to be simply added to the rest, bringing the total uncertainty to \SI{20}{\mega\hertz}.}. 

We report in Fig.~\ref{fig:levelscheme}(d) examples of experimental spectra obtained by scanning the probe-laser frequency at a rate of \SI{5}{\mega\hertz\per\milli\second}. Owing to the short MOT lifetime, this scan speed allows for a continuous measurement of the steady-state atom number. Each detected Rydberg level appears in the spectrum as a pair of twin depletion dips due to the bi-chromatic nature of the MOT light. This feature helps distinguishing Rydberg resonances from other single-dip signals, in general broader and less intense, probably originating from the absorption of two probe photons.

}

\section{Results and discussions}
\label{sec:results}

\begin{figure}[!htbp]
\includegraphics[width=1\linewidth]{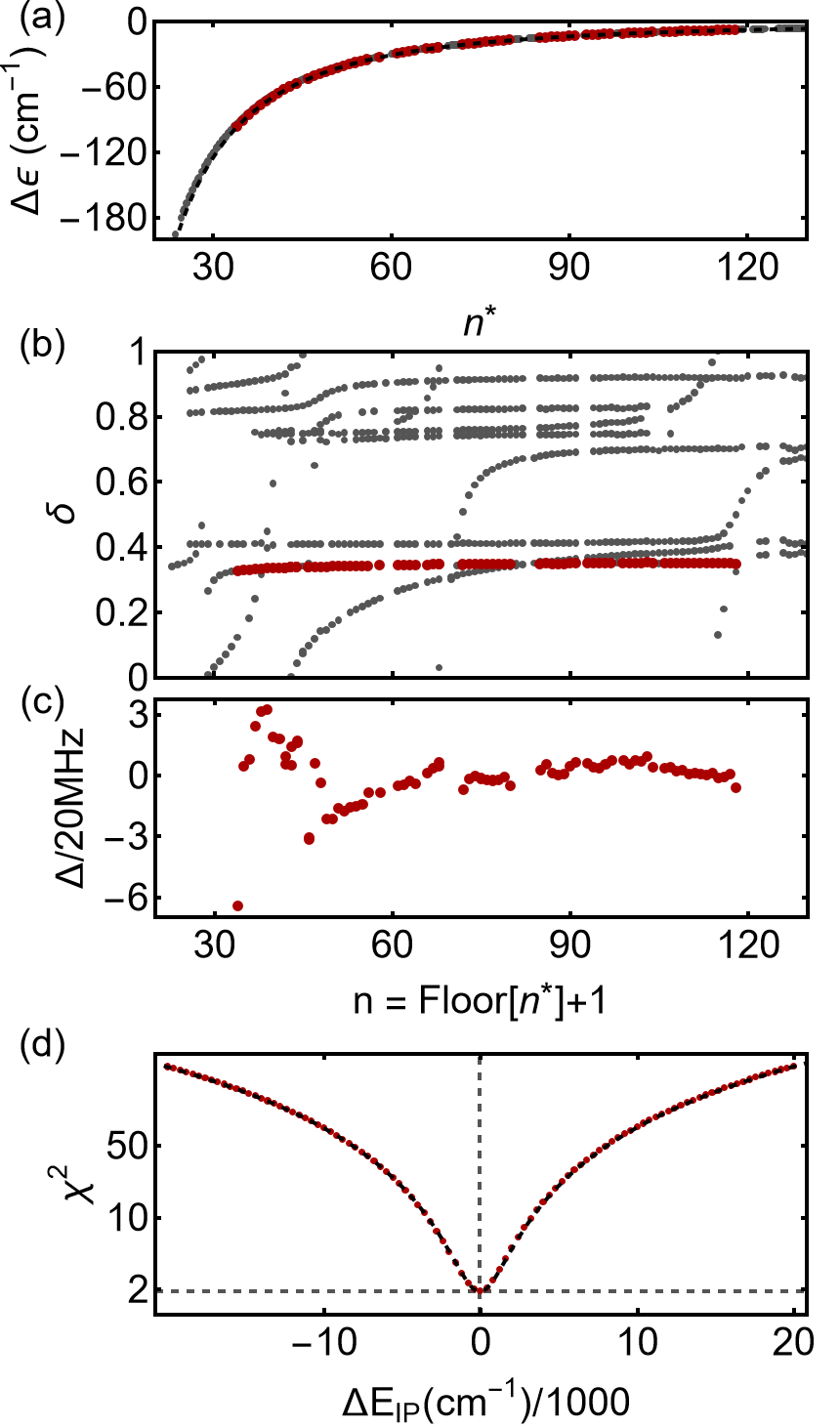}
\caption{{(a) Measured energies of all observed Rydberg levels (b) Plot of the corresponding quantum defects, $\delta$, as a function of the effective integer quantum number $n$. Quantum defects are calculated from Eq.~\eqref{eq:Ryd-Ritz} using the estimated value of  $E_{\mathrm{IP}}$.  For both panels, gray points represent the full experimental dataset, while red points indicate the Rydberg levels used to extract the ionization potential $E_{\mathrm{IP}}$. The black line in (a) is the best fit used to extract $E_{\mathrm{IP}}$.
(c) Residuals ($\Delta$) obtained from the best fit, expressed in units of the experimental uncertainty of \SI{20}{\mega\hertz}
(d) Dependence of the $\chi^2$ value on $E_{\mathrm{IP}}$. The zero of the horizontal axis corresponds to
\SI{47901.8265}{\centi\meter^{-1}}, which is our estimated value of $E_{\mathrm{IP}}$.
The black dashed line shows a parabolic fit to the $\chi^2$ dependence on $E_{\mathrm{IP}}$, used to determine the uncertainty in $E_{\mathrm{IP}}$. {Error bars in (a) and (b) are within the pointsize while in (c) they are 1 by definition, and thus they are not plotted.}
 } }
\label{fig:fig2}
\end{figure}

{In our survey,  the probe laser has been scanned over a spectral range of approximately \SI{250}{\centi\meter}$^{-1}$ below the first ionization threshold, $E_{\mathrm{IP}}$ (wavelength range 
between \SI{414}{\nano\meter} and \SI{419}{\nano\meter}, 
corresponding to the effective principal quantum numbers $n$ ranging from 20 to 130), detecting more than 700 Rydberg levels~\footnote{Data are available from authors upon reasonable request.}. 
}
Fig.\,~\ref{fig:fig2}(a) shows the energies of all observed Rydberg states as a function of the assigned effective principal quantum number $n^*$. The latter is a noninteger quantity defined via the Rydberg-Ritz formula,
\begin{equation}
\label{eq:Ryd-Ritz}
    E_n = E_{\mathrm{IP}} - \frac{\mathrm{R_{162}}}{(n^*)^2}
        = E_{\mathrm{IP}} - \frac{\mathrm{R_{162}}}{\bigl(n - \delta(n)\bigr)^2},
\end{equation}
where $\mathrm{R_{162}}=\SI{109736.9439}{\centi\meter}^{-1}$ is the reduced Rydberg constant for $^{162}$Dy and $E_{\mathrm{IP}}$ denotes the first ionization threshold. From $n^*$, we compute the corresponding effective integer principal quantum number, $n=\mathrm{floor}(n^*)+1$, as well as the quantum defect $\delta(n)=\text{Mod}[-n^*,1]$ associated with a given Rydberg level.

In Fig.~\ref{fig:fig2}(b) we report the quantum defects of the measured levels, illustrating the evolution of $\delta$ with $n$. While the direct representation of Eq.~\eqref{eq:Ryd-Ritz} highlights the expected $1/n^2$ scaling of the binding energy relative to $E_{\mathrm{IP}}$, the energy dependence of quantum defects is particularly useful to resolve distinct Rydberg series, characterized by nearly constant values of $\delta$, and to reveal the presence of perturbers. From our data we identify eight Rydberg series: two series with $0.32 \le\delta\le 0.45$, suggesting an $n{\rm s}$ character~\cite{theodosiou1976electron}; five series with $0.65\le \delta\le  0.95$, consistent with $n{\rm d}$ character; and a final series exhibiting a strong and broad perturbation.

The existence of eight Rydberg series can be anticipated through general angular-momentum coupling arguments and selection rules. Since the intermediate state has total angular momentum $J=9$, optical excitation allows access to Rydberg series with $J=8,9,10$. Furthermore, the ionic core associated with the lowest-energy threshold of Dy$^+$ has character $\rm 4f^{10}(^5I_8)6s(^2S_{1/2})~^6I_{17/2}$. The $n{\rm s}$ Rydberg series converging to this threshold therefore have character $(17/2\; n{\rm s}_{1/2})$ with total $J=8$ or 9. The $n{\rm d}$ series converging to the same threshold can have either $(17/2\; n{\rm d}_{3/2})$ 
\textit{or} $(17/2\; n{\rm d}_{5/2})$, both allowing 
$J=8,9,10$ states. This leads to eight expected Rydberg series (two $n{\rm s}$ and six $n{\rm d}$ orbital angular momentum, or three $J=8$, three $J=9$ and two $J=10$ total angular momentum), in agreement with our observations and assuming the strongly perturbed series to belong to $n{\rm d}$.

\subsection{Determination of the ionization potential}
\label{subsec:IP}

We start our analysis by redetermining the ionization threshold, $E_{\mathrm{IP}}$. 
To extract it, we scan $E_{\mathrm{IP}}$ within a range roughly constrained by previous investigations~\cite{Studer2017dysprosium}.

For each tested value of $E_{\mathrm{IP}}$, we compute $n=\mathrm{floor}(n^*)+1$ and fit some of the observed energies of the $s$ series with $J=8$ (red points in Fig.~\ref{fig:fig2}, and  Sec.~\ref{subsec:angular} for the character definition) using Eq.~\eqref{eq:Ryd-Ritz} together with the truncated expansion for the quantum defect,
\begin{equation}
\label{eq:QuDefsl}
    \delta(n) = \delta_0 + \frac{\delta_2}{(n - \delta_0)^2} + \ldots .
\end{equation}
These experimental points are chosen because they belong to the Rydberg series, among those observed, least affected by perturbers over a wide energy range at high $n$. Moreover, we intentionally exclude two points affected by perturbation ($n =46$ and $47$), as well as the highest-$n$ data, for which residual local electric fields may induce non-negligible energy shifts.

The optimal value of $E_{\mathrm{IP}}$ is obtained by minimizing the fit $\chi^2$, as shown in Fig.~\ref{fig:fig2}(d), yielding a minimum value of $\chi^2 \simeq 1.9$. 
In Fig.~\ref{fig:fig2}(c) we report the best-fit residuals, $\Delta=E_{\mathrm{meas}}(n)-E_{\mathrm{fit}}(n)$, in units of the experimental uncertainty (\SI{20}{\mega\hertz}).

Applying this procedure, we obtain
\begin{equation}
\label{eq:thresh}
     E_{\mathrm{IP}} = 47901.8265 \pm 0.0008\ \mathrm{cm}^{-1},
\end{equation}
which 
{provides a new value of the first ionization limit with an uncertainty reduced by more than one order of magnitude, but  absolute value slightly outside the error bar of the last determination~\cite{Studer2017dysprosium}}. 
The uncertainty on $E_{\mathrm{IP}}$ is evaluated by fitting the computed $\chi^2(E_{\mathrm{IP}})$ with a parabola and extracting the $1\sigma$ confidence interval from the curvature, $\sigma = 1/\sqrt{d\chi^2/dE_{\mathrm{IP}}^2}$. As a cross-check, this fitting procedure was also applied to  $J=9$ $n$s and to $J=10$ $n$d data, producing consistent thresholds but with substantially larger $\chi^2$ value.

From the same fit we obtain $\delta_0=0.3514(2)$ and $\delta_2=-27.4(3)$. 
While $\delta_0$ is in approximate agreement with the value obtained from the MQDT analysis of Sec.~\ref{sec:MQDTfit}, the extracted $\delta_2$ substantially differs from the MQDT result. This discrepancy is expected because Eq.~\eqref{eq:QuDefsl} does not explicitly account for the energy dependence induced by perturbing levels from other Rydberg series. In particular, in this case a perturber located at \SI{143}{\centi\meter}$^{-1}$ below the first ionization threshold (see Sec.~\ref{sec:MQDTfit}) contributes to the apparent energy dependence of $\delta(n)$ and therefore affects the fitted value of $\delta_2$. 

\subsection{Patterns from angular momentum constraints}
\label{subsec:angular}

\begin{figure*}[htbp]
\centering
\includegraphics[width=1\linewidth]{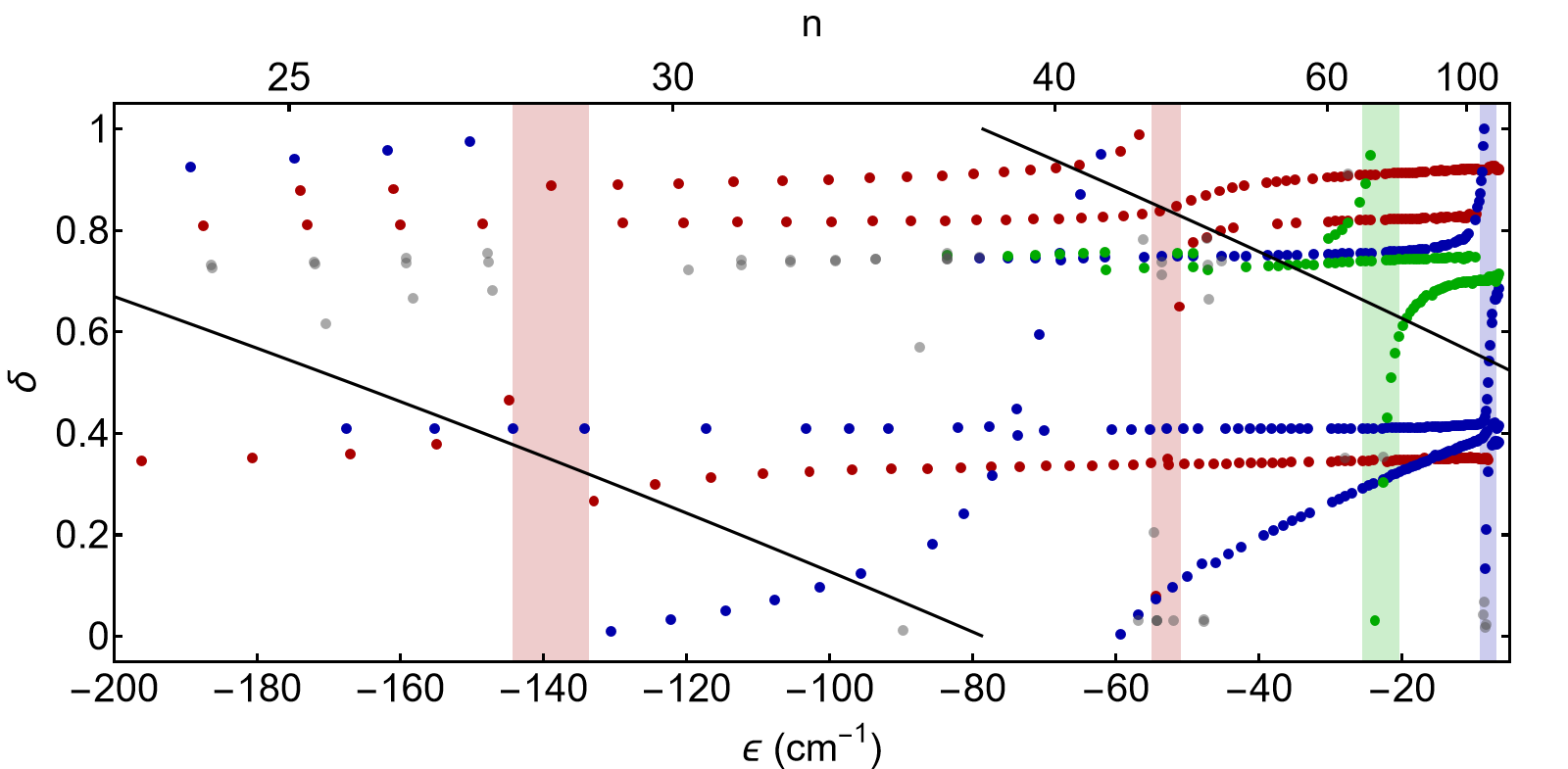}
\caption{Quantum defects of assigned Rydberg levels, divided in series for different values of J. Red points correspond to J=8, blue to J=9 and green to J=10. The colored-shaded areas indicate the position of the perturbers used for series grouping (see main text).  The black line represents the quantum defect $\delta_{15/2}$ calculated from Eq.~(\ref{eq:Ryd-Ritz}) using the first excited ionization threshold;  the discontinuity occurs when $n^{*}_{15/2}$ becomes equal to 11. 
}
\label{fig:fi2b}
\end{figure*}

Although the presence of perturbers complicates the spectrum shown in Fig.~\ref{fig:fig2}(b), these perturbations can be exploited to group and assign the observed Rydberg series. In particular, due to the independence of distinct total angular momentum $J$ manifolds, if a single perturber affects two series, then the two series must belong to the same $J$ manifold. Moreover, the perturbation strength is enhanced when the perturber shares the same orbital character. For instance, an $n{\rm s}$ perturber predominantly affects an $n{\rm s}$ Rydberg series, while its influence on $n{\rm d}$ series is typically weaker.

The first excited state of Dy$^+$ has character $\rm 4f^{10}(^5I_8)6s(^2S_{1/2})~^6I_{15/2}$ and lies \SI{828.314}{\centi\meter}$^{-1}$ above the ionic ground state $\rm 4f^{10}(^5I_8)6s(^2S_{1/2})~^6I_{17/2}$~\cite{Nave2000Dy-levels}. Rydberg states converging to this excited threshold can therefore act as perturbers for series attached to the lower threshold. Given our explored energy range, $\sim250\,\mathrm{cm}^{-1}$, we expect only a small number of such perturbers, corresponding to effective principal quantum numbers in the approximate range $n_{15/2}^*\simeq 10$-$11.5$.

To identify these perturbers, we use the energy relative to the ionization potential,
$\epsilon = E-E_{\rm IP}$, to define the quantity
\begin{equation}
    \delta_{15/2}(\epsilon)=\mod\left[-\sqrt{\frac{\mathrm{R_{162}}}{828.314-\epsilon}},1\right],
\end{equation}
which estimates the quantum defect a perturber converging to the Dy$^+$ $J=15/2$ threshold would have as a function of its energy detuning $\epsilon$ from the first ionization threshold $E_{\mathrm{IP}}$. This estimate is shown as  black lines in Fig.~\ref{fig:fi2b}.

Because the ionic core has $J=15/2$, the corresponding $n{\rm s}$ series can only have total angular momentum $J=7$ or $J=8$. Therefore, the observation of a perturber predominantly affecting an $n{\rm s}$ series strongly suggests that the perturbed series belongs to the $J=8$ manifold. This is precisely the situation for the perturbation located at $\epsilon \simeq \SI{-143}{\centi\meter}^{-1}$, highlighted by the red-shaded region in Fig.~\ref{fig:fi2b}. In this region $\delta_{15/2}\simeq 0.38$, consistent with an $n{\rm s}$ perturber, supporting an interpretation in terms of coupling to an $n{\rm s}$ series converging to the Dy$^+$ excited threshold. This argument unambiguously assigns the series with unperturbed quantum defect $\delta\simeq 0.35$ to the $J=8$ manifold.

The identification of the remaining $J=8$ series can be completed by noting that the two Rydberg series with unperturbed quantum defects $\delta\simeq 0.90$ and $\delta\simeq 0.82$ share a common perturbation at $\epsilon\simeq \SI{-52}{\centi\meter}^{-1}$ with the above-described $J=8$ $n{\rm s}$ series. For this second perturber, $\delta_{15/2}\simeq 0.84$, suggesting an $n{\rm d}$ character. This also explains the stronger perturbation observed on the two $n{\rm d}$ series compared to the $n{\rm s}$ series. Beyond identifying the $J=8$ manifold, the reasonable values of $\delta_{15/2}$ suggest that both perturbers belong to Rydberg series converging to the first excited Dy$^+$ threshold.

Using similar arguments, we note from Fig.~\ref{fig:fi2b} that the perturber located at $\epsilon\simeq \SI{-8}{\centi\meter}^{-1}$ affects three different series. This implies that these three series form the $J=9$ manifold. Among them, the series with unperturbed quantum defect $\delta\simeq 0.41$ is assigned to $n{\rm s}$ character, while the other two are $n{\rm d}$ series, including the one exhibiting a broad perturbation centered around $\epsilon\simeq \SI{-70}{\centi\meter}^{-1}$.

Finally, the two remaining series must correspond to the $J=10$ manifold, since they share a common perturber at $\epsilon\simeq \SI{-22}{\centi\meter}^{-1}$. Their unperturbed quantum defect $\delta\simeq 0.73$ is consistent with the expected values for $n{\rm d}$ character.

{While the analysis based on constraints imposed by angular momentum was able to assign many of the observed levels, for others this approach is too simplistic. This is the case, for instance, for the points with $n$ between 35 and 90 and $\delta\simeq 0.72$, where many different lines have similar quantum defects. To refine the assignment and to study the effect of the observed perturbers more quantitatively, in the next Section, we fit our data with a simplified parameterization of Multi-Channel Quantum Defect theory (MQDT).}

\section{MQDT approach}
\label{sec:MQDT}


In this Section, we describe how we use ideas from 
MQDT~\cite{Aymar1996MQDT} to understand some properties of the observed Rydberg states. In particular, we show 
{ how perturbers affect the series MQDT parameters, giving a
quantitative description of the observed levels at near experimental uncertainty. We fit the
line positions over a range of $\sim {250}\,{\rm cm}^{-1}$ (approximately \SI{7.5}{\tera\hertz}) with
a standard deviation of approximately 150, 110, and \SI{70}{\mega\hertz} for the $J=8$, 9, and 10 series, respectively.
The goal is to use the MQDT fit to classify lines within each series and to identify
Rydberg states that are strongly modified by perturbing states attached to higher
thresholds.} 
Moreover, we use ideas
from the frame transformation approximation to understand the energy of the $J=8$ $n$s perturber.

We  use the representation where a real, symmetric $K$-matrix gives the
coupling between the different channels, i.e. between the different Rydberg series. 
The equations of MQDT
for a {\it single} threshold, as we are modeling for Dy,
show that the bound states are determined by
\begin{equation}\label{eq:MQDTbnd}
\det [\mathbf{K}(\epsilon ) +\tan\pi\nu (\epsilon )\mathbf{I}]=0
\end{equation}
where $\epsilon = E-E_{\rm IP}$ is the energy in cm$^{-1}$ {\it relative to the threshold},
$\mathbf{I}$ is the unit matrix, and $\nu (\epsilon )=\sqrt{-R_{162}/\epsilon}$. 
Note that: 1) $\nu(\epsilon)$ is defined for any negative energy value, $\epsilon$, while $n^*$, which shares the same definition, has value only for the actual bound states; 2) any unitary transformation of the $K$-matrix,
$\tilde{\mathbf{K}}=\mathbf{UKU}^\dagger$ with $\mathbf{U}$ a unitary matrix,
will still satisfy Eq.~(\ref{eq:MQDTbnd}) so a fit of $\mathbf{K}$
typically only constrains the eigenvalues.

When the perturbers are separated in energy, the $K$-matrix can be
approximated by the form
\begin{equation}\label{eq:MQDTK}
K_{ij}(\epsilon )=\delta_{ij}\tan\pi\delta_i(\epsilon )-
\frac{V_{i,1}V_{j,1}}{\epsilon -E_1}-
\frac{V_{i,2}V_{j,2}}{\epsilon -E_2}-...
\end{equation}
where the quantum defects $\delta_i(\epsilon )$ slowly depend on energy,
$E_\alpha$ is the energy of the $\alpha$-th perturber, and the
$V_{i,\alpha}$ are real numbers that parameterize the interaction of the
$\alpha$-th perturber with channel $i$. Given the fact that series with different angular momentum $J$ do not interact in the absence of
external fields, channels with different $J$ are treated separately.
For one channel and one perturber,
the energy width of the interaction is $\Gamma = 2 V^2/(1 +\tan^2\pi\delta )$. For
more channels, the relations are more complicated, but, roughly, larger
$|V_{i,\alpha}|$ means perturber $\alpha$ has a larger effect on channel $i$.
A sketch of the derivation of this form is given in
App.~\ref{sec:mqdtpert}. Because the fit does not
constrain a unitary transformation of $\mathbf{K}$,
we have chosen to have the slowly varying portion of Eq.~(\ref{eq:MQDTK})
be diagonal. The $\delta_i(\epsilon)$ can have the 
form of Eq.~(\ref{eq:QuDefsl}) or an equivalent, easier to implement form:
\begin{equation}\label{eq:MQDTsl}
 \delta_i(\epsilon)=\delta_{0,i}+\epsilon\frac{d \delta_i (\epsilon)}{d\epsilon}\vert_{\epsilon =0}   
\end{equation}
where the $\delta_{0,i}$ and the derivative are evaluated at threshold. This form is related
to that in Eq.~(\ref{eq:QuDefsl}) 
through $\delta_{2,i} =-R_{162} d \delta_i (\epsilon)/d\epsilon$.
The difference between the two forms is $\delta_2^2/\nu^5$ which is negligible
for our states and accuracy.

\begin{figure}[!htbp]
\centering
\includegraphics[width=1\columnwidth]{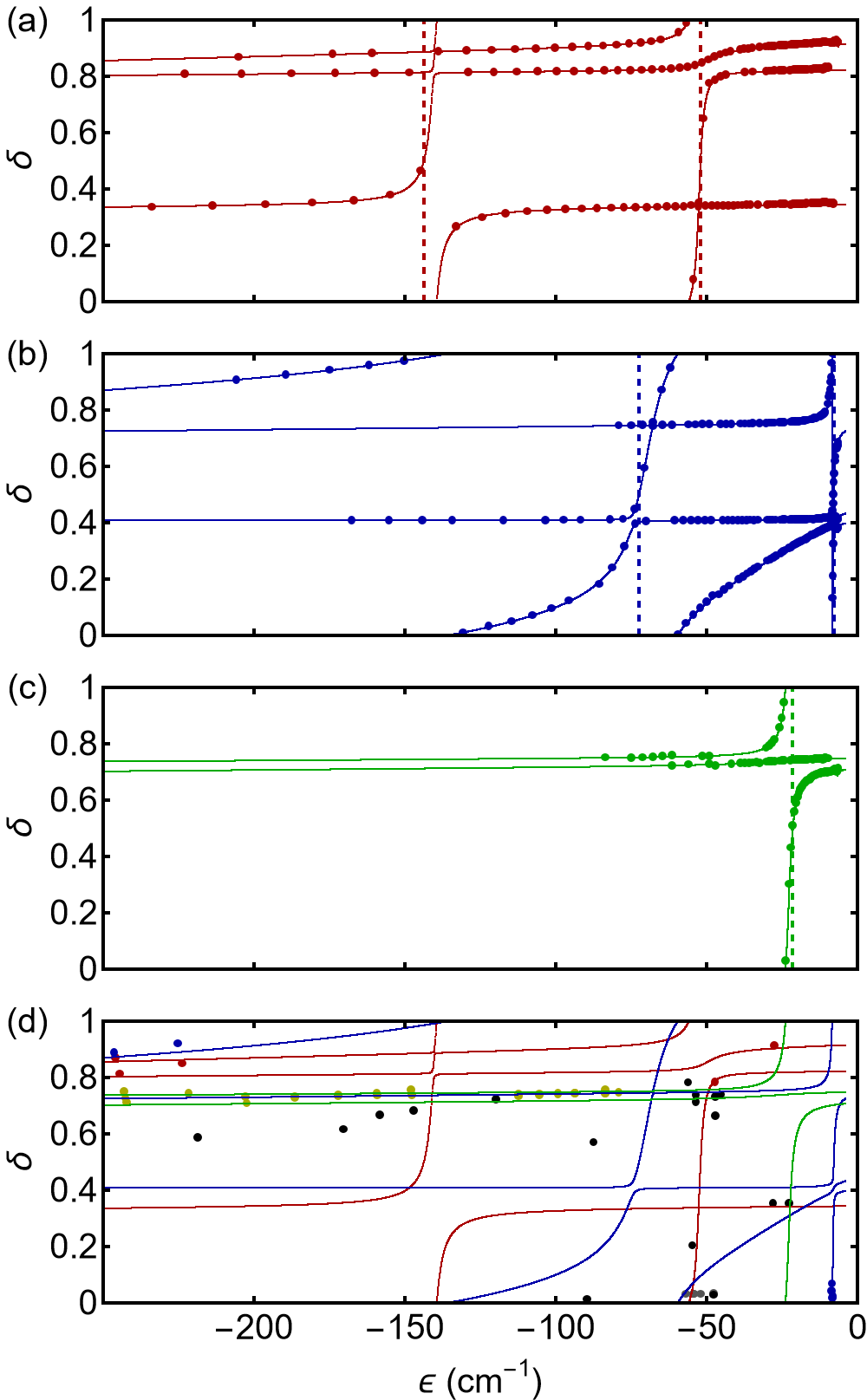}
\caption{{Panels (a), (b), and (c) show the results of MQDT calculations compared with the experimental data for the Rydberg series with $J=8$, 9, and 10, respectively. The fitting routine is also used to assign a total angular momentum $J$ to each experimental data point. The dashed vertical lines represent the position of the perturbers reported in Tab.~\ref{tab:perturbers}. The $J=9$ case possesses a perturber not shown in the Figure, located above $E_{\mathrm{IP}}$.
(d) Experimental data points that remain unassigned. Red and blue points indicate points suspected to correspond to J=8 and 9, respectively, 
because they are quite near a  quantum defect line from the fit, but a physical constraint disqualifies it for that $J$. 
{Yellow points represent experimental data having either J=9 or J=10.} Gray points may belong to a Rydberg g-series, as suggested by their small quantum defect. Black points are not assigned either because theoretical predictions for different series are nearly indistinguishable or because the experimental values deviate too strongly from the theoretical expectations.
}}
\label{fig:defects}
\end{figure}

The results of the fitting are shown in Fig.~\ref{fig:defects}, where the lines are obtained by taking
the arctangent of
the eigenvalues of the energy dependent $K$-matrix at each energy then dividing
by $\pi$; we use the definition of arctangent so the result is between 0 and 1. 
When the energy exactly equals one of the perturber's energies,
$\epsilon = E_\alpha$, then one of the eigen-quantum defects equals $1/2$.

\subsection{Results of MQDT fit}
\label{sec:MQDTfit}

\begin{table}[htb]
    \centering
    \begin{tabular}{|c|c|c|}
        \hline
        \textbf{} & \textbf{$\delta_{0,i}$} & \textbf{$\delta_{2,i}$}  \\
        \hline
        \hline

        \multirow{3}{*}{J=8} &\null\quad 0.34710(17)\null\quad\null &\null\quad -7.73(12)\null\quad\null \\
                                  & 0.82402(23) & -9.65(19)  \\
                                  & 0.92174(24) & -29.87(19)\\
        \hline
        \hline

        \multirow{3}{*}{J=9} & 0.40740(26) & -0.09(21) \\
                                  & 0.88622(20) & -37.48(15)\\
                                  & 0.74936(65) & -11.42(1.2)\\
        \hline
        \hline

        \multirow{3}{*}{J=10} & 0.72236(66) & -9.4(1.7) \\
                                  & 0.75275(48) & -7.43
                                  (80) \\
                                 
        \hline
    \end{tabular}
    \caption{MQDT fit with Eqs.~(\ref{eq:QuDefsl}) and (\ref{eq:MQDTsl}): fit for the quantum defects $\delta_0$ and $\delta_2$ for the eight observed series. The parenthesis gives the parameter {{\it width} from the fit. See App.~\ref{sec:MQDTerrors} for a discussion of the uncertainty in the MQDT fit parameters.}
    }
\label{tab:unperturbed_QD}
\end{table}

\begin{table}[htb]
    \centering
    \begin{tabular}{|c|c|c|c|c|}
        \hline
        \textbf{} & \textbf{$E_\alpha$} cm$^{-1}$ & \textbf{$V_{1,\alpha}$} cm$^{-1/2}$& \textbf{$V_{2,\alpha}$}cm$^{-1/2}$ & \textbf{$V_{3,\alpha}$}cm$^{-1/2}$ \\
        \hline
        \hline

        \multirow{3}{*}{J=8} & -52.051(16)  & 0.002(43) & 0.641(13) & 1.0016(74) \\
        & -143.696(30)  & 2.7674(86) & 0.146(69) & 0.057(85) \\
                         
        \hline
        \hline

        \multirow{3}{*}{J=9}  & +9.95(14)  & 1.22(13) & 8.0639(64) & 1.208(63) \\
        & -7.939(39)  & 0.606(93) & -0.14(18) & 0.818(27) \\
        & -72.453(26) & 0.670(40) & 2.5305(52) & 0.150(20) \\
                                 
        \hline
        \hline

       J=10  & -21.5292(94)  & 1.3547(57)  & 0.9425(65) &  \\
                                 
        \hline
    \end{tabular}
    \caption{MQDT fit with Eq.~(\ref{eq:MQDTK}) for perturber parameters:  The parenthesis gives the parameter {{\it width} from the fit. See App.~\ref{sec:MQDTerrors} for a discussion of the uncertainty in the MQDT fit parameters.} In Ref.~\cite{Studer2017dysprosium}, the perturber at \SI{-143.70}{\centi\meter}$^{-1}$ in the $n$s series had already been reported, and the strongly perturbed series corresponding to the level at \SI{+9.95}{\centi\meter}$^{-1}$ had been observed but not interpreted.}
\label{tab:perturbers}
\end{table}

We fixed the threshold to the value in Eq.~(\ref{eq:thresh}), allowed
all parameters defining the $K$-matrix, Eq.~(\ref{eq:MQDTK}), to vary and fit these parameters by minimizing the reduced $\chi^2$. 
The number of fit parameters is

\begin{equation}
N_{fit} = 2 N_{ch} + N_{pert}\times (N_{ch}+1)
\end{equation}
where $N_{ch}$ is the number of channels and $N_{pert}$ is the number of perturbers for
that $J$.
The first term in $N_{fit}$ is from Eq.~(\ref{eq:MQDTsl}) and the last term is from
the perturbers. We use these parameters for the series to fit 245 $(J=8)$,
256 ($J=9)$, and 119 $(J=10)$
energies using 14 $(J=8)$, 18 ($J=9)$, and 7 $(J=10)$ parameters.
The minimum values for the $\chi^2$ were 57.2, 29.2, and 11.8 for
$J=8$, 9 and 10 respectively. These large values of $\chi^2$ are due to an oversimplified modelling, with a reduced number of perturbers and couplings~\footnote{There are two aspects of particular difficulty. The first is that there are many lines
below \SI{-85}{\centi\meter}$^{-1}$ which could not be definitively classified but
belong to either $J=9$ or $J=10$. Below this energy, there is a near degeneracy of one of the
quantum defects of $J=9$ with that of $J=10$, at the value $\delta\simeq 0.74$, leading to pairs of
close levels. The second main difficulty is that of the signs
of the $V_{i,\alpha}$ are not well defined. Arbitrarily, we set the values of
the $V_{i,\alpha}$ to be positive when they are not consistent with zero.}.

For each of the 8 Rydberg series identified, the $\delta_{0,i}$ and
$\delta_{2,i}$ are reported in Tab.~\ref{tab:unperturbed_QD}.
As a point of comparison, the $\tau_0,\tau_2$ in Table~II of Ref.~\cite{trautmann2021spectroscopy}
(corresponding to our $\delta_0,\delta_2$) are $0.301,-15.78$
and $0.427,-1.718$ for the $n$s series in Er
compared with $0.347,-7.7$ and $0.407,\sim 0$ for our fits to the Dy $n$s series. The values
are not expected to be the same since the specific values depends on details of
angular momentum coupling, but they should be similar.

In Tab.~\ref{tab:perturbers},
we report the values for all the perturbers identified from the MQDT fit. Each perturber is also associated to 2 or 3 coupling coefficients, $V_{i, \alpha}$, depending on the value of $J$ of the perturbing state. The discussion of the uncertainty in the fitting parameters is in App.~\ref{sec:MQDTerrors}.
{
From the MQDT fits we obtained precise estimates of the perturber energies, which can be used to gain insight into their physical nature.

As already discussed in Sec.\,\ref{subsec:angular}, most of the observed perturbers can be confidently identified as Rydberg levels attached to the first excited state of Dy$^+$. In particular, for the perturbers affecting the $J=8$ manifold at \SI{-143.696}{\centi\meter}$^{-1}$ and \SI{-52.050}{\centi\meter}$^{-1}$; the $J=10$ perturber observed at \SI{-21.527}{\centi\meter}$^{-1}$; and two of the $J=9$ perturbers with negative energies at \SI{-72.451}{\centi\meter}$^{-1}$ and \SI{-7.932}{\centi\meter}$^{-1}$, we can evaluate the effective quantum number $n^*_{15/2}$ and the corresponding quantum defect $\delta_{15/2}$ as defined in Sec.\,\ref{subsec:angular}. This yields the pairs $(10.625,\,0.375)$, $(11.165,\,0.835)$, $(11.363,\,0.637)$, $(11.037,\,0.963)$, and $(11.455,\,0.545)$, respectively.

All of these values correspond to plausible quantum defects for $n$s (the first case) and $n$d (the remaining cases) Rydberg series. The only apparently anomalous value is the quantum defect obtained for the $J=9$ perturber at \SI{-7.932}{\centi\meter}$^{-1}$. However, the value $\delta_{15/2}=0.545$ can be readily explained by assuming values of the channel-dependent defects $\delta_i$ similar to those extracted from our fits. For instance, taking $\delta_0$ to be a small $n$d value from Table~\ref{tab:unperturbed_QD} (e.g., $\delta_0=0.72$) and $\delta_2$ to have a moderate magnitude (e.g., $\delta_2=-20$), one obtains a total quantum defect
$\delta_{15/2} = 0.72 - \frac{20}{11.28^2} \approx 0.56$,
which is fully consistent with the fitted value. This supports the interpretation of this perturber as having predominantly $15/2\,n$d character.

The final perturber to consider is the positive-energy $J=9$ state observed at \SI{9.95}{\centi\meter}$^{-1}$. Applying the same procedure yields $(n^*_{15/2},\delta_{15/2}) = (11.581,\,0.419)$, which would nominally suggest an $n$s character. However, the $J=9$ manifold with $15/2$ character does not support an $s$-series. The exceptionally strong coupling of this level to the other fitted Rydberg series instead leads us to interpret this perturber as an $n$d Rydberg level attached to the second excited ionic state, $\text{4f}^{10}({}^5I_7)\,6s({}^2S_{1/2}),{}^6I_{15/2}$,
which lies \SI{4341.104}{\centi\meter}$^{-1}$ above the first ionization threshold. Under this assignment, the corresponding quantum defect is $\delta \approx 0.966$, with an effective quantum number $n^* \approx 5.034$. The small value of $n^*$ would imply that the perturber is strongly bound, which could also explain the large coupling strength observed. }

{Finally, we also stress that MQDT fits have also been used to spot the rare misidentification ($\ll\,1\%$ of the total) within the experimental data. They were mainly originated by mode jumps or instabilities of the probe laser.}

\subsection{Frame transformation for ns}\label{sec:FT}

For many atoms with $n$s and $n$d Rydberg states, perturbers interact most strongly
with series with the same outer angular momentum. Thus,
$n$s series tend to weakly interact with $n$d perturbers attached to higher thresholds
and $n$d series tend to weakly interact with $n$s perturbers. As discussed in the
previous section, this suggests that the
perturber at $\SI{-143.696}{\centi\meter}^{-1}$ has $n$s character attached to the 15/2 threshold.
This was the main clue for distinguishing the $J=8$ and 9 series since only $J=8$ 
has $n$s Rydberg series attached to the 15/2 threshold.

To make this assignment more firm, we use ideas from the frame transformation
approximation, Sec.~IIE of Ref.~\cite{Aymar1996MQDT}, for the two $n$s series and the
$n$s perturber. The basic idea behind the frame transformation is that $LS$ coupling is a good
approximation when the Rydberg electron is at small distances. This gives a phase shift
that depends on whether the 6s and $n$s electrons are coupled to singlet or triplet, but
it does not couple the two even though they have the same total $J$ after coupling with
the 4f electrons. However, when the Rydberg electron is more than $\sim 20$~$a_0$ beyond
the core electrons, it is more appropriate for the 6s to be coupled to the 4f electrons to
give a $J$ for the core and
then the Rydberg electron's angular momentum is coupled to the core angular momentum. 
These two regions can be connected through a unitary transformation which arises from
the different order of angular momentum coupling. The main point is that this approximation
gives 4 $K$-matrix elements in terms of 2 phase shifts. If the data from the two $n$s series and
one $n$s perturber can be represented this way, that gives a strong indication that the
identifications of the $J$ for the different series are correct.

Appendix~\ref{sec:ftdetail} gives the details of the frame transformation which leads to
4 $K$-matrix parameters given in terms of a specified frame transformation matrix and
two unspecified diagonal elements, Eq.~(\ref{eq:ftappr}). From Eq.~(\ref{eq:deriv})
with $E_{15/2}-E_\alpha = \SI{972.01}{\centi\meter}^{-1}$ and the $\nu_\alpha =10.625$, allows the
calculation of all 4 matrix elements from experimental data:
\begin{eqnarray}
K^{(9)}_{17/2,17/2}&=&\tan\pi 0.4074=3.340\nonumber\\
K^{(8)}_{17/2,17/2}&=&\tan\pi 0.3470=1.918\nonumber\\
K^{(8)}_{15/2,15/2}&=&\tan\pi 0.375=2.414\nonumber\\
K^{(8)}_{15/2,17/2}&=&V\sqrt{\frac{d\tan\pi\nu}{d\epsilon}}=0.947
\end{eqnarray}
where we used $V= \SI{2.767}{\centi\meter}^{-1/2}$. We used these four values and the form
Eq.~(\ref{eq:ftappr}) in a $\chi^2$
to find the best values for the two unknowns. The $\chi^2$ minimization
gave $K_{^3S}=3.259=\tan (0.405\;\pi)$ and $K_{^1S}=1.154=\tan (0.273\;\pi)$. By using these values in Eq.~(\ref{eq:ftappr}), we can compare MQDT fitting and the frame transformation approximation. Results are shown in Tab.~\ref{tab:FT-MQDTcomparison}, where the rightmost values are from Eq.~(\ref{eq:ftappr}).

\begin{table}[htb]
    \centering
    \begin{tabular}{|c|c|c|}
        \hline
        \textbf{K-matrix parameter} & \textbf{MQDT fit} & \textbf{FT approximation}  \\
        
        \hline
        \hline

        $K_{17/2,17/2}^{(9)}$  & 3.340  & 3.259 \\
        
        \hline

        $K_{17/2,17/2}^{(8)}$  & 1.918  & 2.145 \\
        
        \hline

        $K_{15/2,15/2}^{(8)}$  & 2.414  & 2.268 \\      
        
        \hline

        $K_{15/2,17/2}^{(8)}$  & 0.947  & 1.051 \\
        
        \hline
    \end{tabular}
    \caption{Comparison between MQDT fitting and frame transformation approximation for the two $n$s Rydberg series.}
    
\label{tab:FT-MQDTcomparison}
\end{table}


This is reasonable agreement for such a simple approximation. Remember, we are using the same quantum
defect whether it is relative to the 17/2 or 15/2 threshold while the fits to the Rydberg
series attached to the 17/2 thresholds suggest the values for
$\delta_2$ are not 0. The largest difference in the diagonal elements is for the
$K^{(8)}_{17/2,17/2}$ and corresponds to a difference in quantum defect of 0.014.

A similar transformation could be performed for the $n$d series. However, we did not attempt
this study since these transformations will
lead to $4\times 4$ matrices and our fit data do not give all of the relevant parameters. 


\section{Intensity of depletion lines}
\label{sec:amplitude}

Most applications of Rydberg states, either implying resonant Rabi flopping or off-resonance dressing, require a thorough characterization of the spectrum, integrating the tabulation of the resonance frequencies with additional information on the oscillator strength and  lifetime of the excited state. In this respect, s%
ince our spectroscopic investigation is performed at fixed laser intensity, applying a naive approach one might expect 
to extract the oscillator strength from the amplitude of the depletion signal.
We find that, at least in the clearest case of $J=10$, 
such an analysis is able to reveal the peaked modulation in the presence of a pertuber, originating from the coupling with a state of lower principal quantum number. 
\begin{figure}
    \centering
    \includegraphics[width=1\columnwidth]{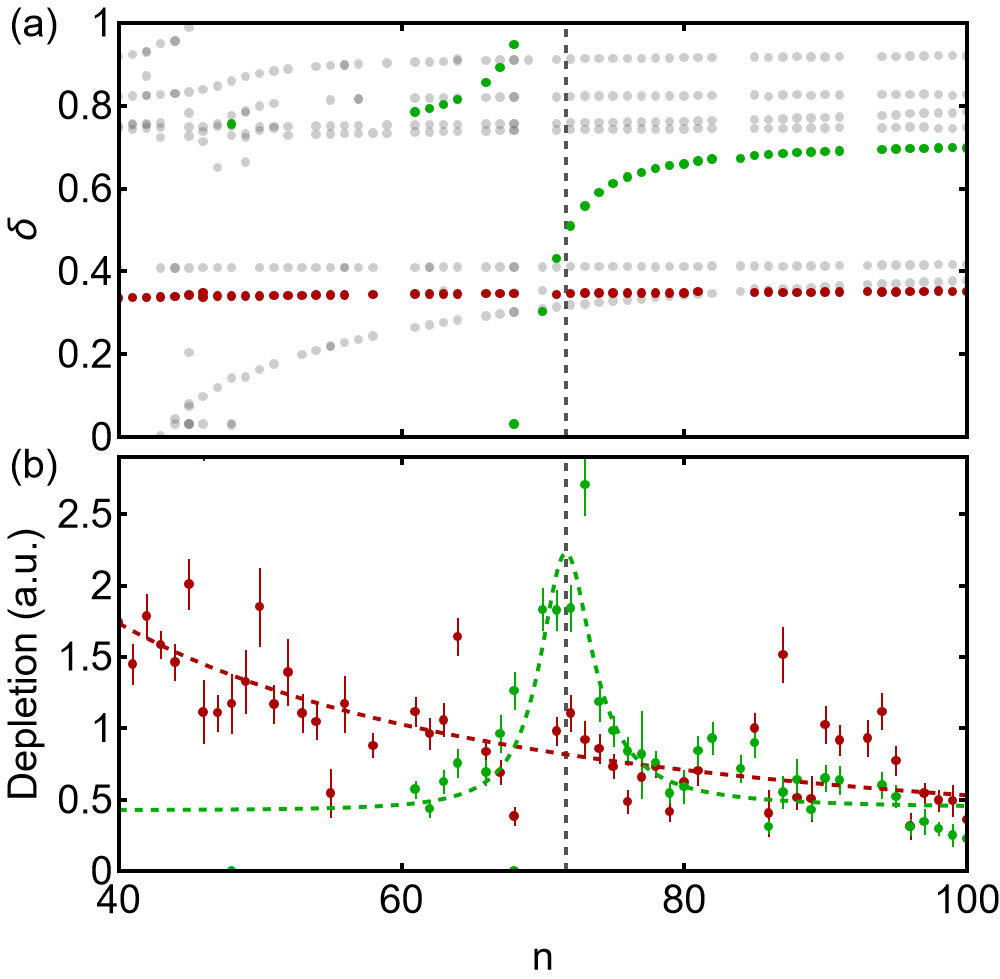}
\caption{In (a) quantum defects of all the 8 observed Rydberg series as a function of $n$. In red the unperturbed $J=8$, and in green the perturbed $J=10$ Rydberg series reported in the next panel. In (b) amplitude of the depletion signals for the two $J=8$ and $J=10$ Rydberg series. Dashed lines are best fit functions (see text).}
    \label{fig:fig4}
\end{figure}
Fig.~\ref{fig:fig4} 
shows the experimental measurements of the depletion signal (defined as the product of the signal amplitude and its full width at half maximum) associated with the strongly perturbed $J=10$ series and the least perturbed $n$s $J=8$ series (the one selected for the estimation of $E_{\mathrm{IP}}$). 
In the former case, the trend exhibits a pronounced resonance around $n \simeq 72$, whereas in the latter it decreases monotonically with increasing principal quantum number.
By assuming a Lorentzian behavior with coupling $\Gamma$ in the energy domain, we fit the experimental data obtained in presence of a perturber using the following phenomenological function:


\begin{equation}
f(n^*)=\alpha\frac{1}{1+(\epsilon(n^*)-\epsilon_0)^2/(\Gamma/2)^2}
\label{eq:fit}
\end{equation}
where $\alpha$ is the amplitude of the feature in the depletion signal, and $\epsilon_0$ and $\Gamma$ denote the energy detuning and the coupling strength
between the Rydberg series and the perturber, respectively. $\epsilon(n^*)$ and $\epsilon_0$ represent the detuning as determined by  Eq.~(\ref{eq:Ryd-Ritz}) and the energy position of the perturber, respectively, both {relative to the first ionization limit}. 
From the fit, we extract a resonance position at $\SI{-21.4 \pm 0.2}{\centi\meter}^{-1}$ and a coupling strength $\Gamma = \SI{2.8 \pm 0.6}{\centi\meter}^{-1}$. While the position of the perturber is in agreement with the MQDT calculations reported in Tab.~\ref{tab:perturbers}, its width is only compatible.
Conversely, for the unperturbed series, the data are fitted with a power-law dependence $n^{-\beta}$, yielding $\beta = 1.3 \pm 0.3$. Although this behavior clearly indicates a monotonic decrease of the depletion signal with increasing $n$, the fit results show that the oscillator strength alone is insufficient to fully describe the observed trends, as it would determine a $1/n^3$ dependence.

In fact, the dependence of the MOT loss rate on the Rydberg excitation is determined by the combination of a number of processes~\cite{Walker2008traploss, Raithel2001high-l} with different trends over $n$~\footnote{The excitation from ground-state to a Rydberg level may results in trap-loss due to a combination of effects, as discussed in ref.~\cite{Walker2008traploss} for the Rb case. An atom excited to a Rydberg level can exit the cooling/trapping cycle because it can, with some probability, decay back into some metastable level and fall outside the trapping region before going back to the real ground state. Alternatively, the Rydberg atom can migrate into neighboring higher-$l$ levels with longer radiative lifetimes~\cite{Raithel2001high-l} and again leave the trap before being recycled, or be ionized by either a black-body or a laser photon. Other processes like Rydberg-Rydberg collisions may also be possible but are supposed to play a minor role.}. Moreover, the interpretation of the signal is further complicated as we probe the absolute population in the intermediate state of a ladder three-level system with incoherent constant inflow and variable losses. For these reasons, we do not expect the depletion signal to provide a direct or quantitative measure of the oscillator strength.

Despite these limitations, the depletion observable allows us to clearly distinguish qualitative differences induced by the presence of a perturber. Remarkably, a quantitative analysis of the $J=10$ series yields results consistent with MQDT predictions, thereby providing an independent benchmark of the theoretical model.
A more comprehensive comparison between the experimental spectra (including the other lines) and the oscillator strengths evaluated within MQDT would require a detailed and quantitative characterization of all the physical mechanisms responsible for the depletion, which goes beyond the scope of the present work.

\section{Conclusions and outlook}
\label{sec:conclusions}
In conclusion, we present the first high-resolution Rydberg spectroscopy of $^{162}$Dy, realized by observing trap depletion in a magneto-optical trap. We classify more than 600 levels (over roughly 700 detected ones), assigning them to the 8 expected Rydberg series we can explore with our excitation scheme. Additionally, our measurements yield an improved estimate of the first ionization energy, refined by an order of magnitude compared to previous studies. The experimental data are supported by MQDT fits. Besides confirming and completing the identification of the Rydberg series, MQDT fits have been used to estimate the position and the coupling of the six observed perturbers within the explored energy region. Finally, by analyzing the amplitude of the depletion signal, we qualitatively observed how a perturber enhances the coupling to the Rydberg series it is affecting.

From a spectroscopic standpoint, a natural extension of this work would be to conduct high-resolution spectroscopy in the presence of magnetic~\cite{trautmann2021spectroscopy} or electric fields, in order to further validate the line assignments presented here. This would also allow for the precise determination of the polarizability of the Rydberg states.


Although our present study deals with the excitation of low-$n$ Rydberg states from the external 6s shell, many intriguing possibilities may arise from the simultaneous excitation of an internal isolated-core 4f electron. Fast auto-ionization of these states could be mitigated as in the Yb case~\cite{Cheinet2022LightshiftfromICE}, giving access to optical trapping of Rydberg atoms, eventually with high-$l$ excitation~\cite{Ravon2023CircularRydberg}, and double-Rydberg studies~\cite{Camus1989Observation}.

\begin{acknowledgments}
We acknowledge support from 
the European Union through the ERC SUPERSOLIDS project n.101055319, and the QuantERA Programme, project MAQS, under Grant Agreement n.101017733, with funding organisation Consiglio Nazionale delle Ricerche. We acknowledge support from the European Union - NextGenerationEU for  PNRR MUR Project 'National Quantum Science and Technology
Institute - NQSTI' (Partenariato esteso 04: Scienze e Tecnologie
Quantistiche - PE 0023, CUP B53C22004180005).
A.F. and L.T. acknowledge funding from the Italian MUR (PRIN DiQut Grant No. 2022523NA7)
FR was supported by the National Science Foundation under Award No. 2410890-PHY.
We thank Luca Tanzi, Giulia Semeghini, Igor Ferrier-Barbut, Antoine Browaeys, Steven Lepoutre, Patrick Cheinet, Daniel Comparat and Thomas Gallagher for very fruitful discussions. We thank Leonardo Salvi and his team for providing the high resolution wavemeter and its calibration.
\end{acknowledgments}

\appendix

\section{Uncertainty in MQDT fit parameters}\label{sec:MQDTerrors}

To obtain an estimate of the uncertainty in the fit parameters, we do not use the full variance
matrix. We vary one parameter at a time.
For a given parameter $Q$, one-half the separation of the two values that give $\chi^2_{min}+1/2$
defines $\Delta Q$.
For $J=10$, there is only one perturber so the signs of $V$ can not
be found from the fitting procedure. For the other $J$, we found that we could change the sign of
the $V_{i,\alpha}$ and get nearly as good fits but with some other parameters changing
in size, some of them beyond the listed values of their width.
As an example, changing the \SI{0.641}{\centi\meter}$^{-1/2}$ for the $J=8$ resonance at
\SI{-52.051}{\centi\meter}$^{-1}$
to minus value left the $\chi^2$ almost unchanged and hardly changed any of the other
parameters; clearly, the uncertainty in this parameter is not the \SI{0.013}{\centi\meter}$^{-1/2}$ listed in the table since
either value is acceptable. As another example for $J=9$, when we set the
$V_{1,\alpha}=\SI{1.22}{\centi\meter}^{-1/2}$ for the $\SI{9.95}{\centi\meter}^{-1}$ perturber to be negative,
the minimization gave the other coupling parameters outside
the range from the original $\chi^2$ fit by factors less than 2.

\section{Derivation of $K$-matrix near perturbers}\label{sec:mqdtpert}

In this Section, we give a brief derivation of the approximation of Eq.~(\ref{eq:MQDTK}).
To simplify the derivation, we will only consider two thresholds. The higher
energy threshold leads to perturbers for the lower thresholds. In the sense of MQDT,
the $K$-matrix including both thresholds hardly has energy dependence. Channels attached to the
lower threshold will get subscript $r$ (for Rydberg) and those attached to the upper threshold
will get subscript $p$ (for perturber). We can perform a
unitary transformation on the $K$-matrix so that the part of the $K$-matrix in the perturber
space and the part in the Rydberg space are each diagonal.
Applying the bound state boundary conditions
to the perturber channels~\cite{Aymar1996MQDT} gives the $K$-matrix
\begin{equation}\label{eq:MQDTclo}
K_{r,r'}=\delta_{r,r'}\tan\pi\delta_{r}-\sum_{p}\frac{K_{r,p}K_{p,r'}}{\tan\pi\nu_p + \tan\pi\delta_p}
\end{equation}
where the $K_{r,r'}$ is the $K$-matrix in the Rydberg space after applying the boundary conditions
to the perturbers, the $\tan\pi\delta_{r}$ and $\tan\pi\delta_{p}$ are the diagonal elements of the
$K$-matrix in the Rydberg and perturber space respectively before
applying the boundary conditions, and the $\nu_p=\sqrt{R_{162}/(E_p-E)}$
with $E_p$ the energy of the perturber threshold. This form does not have approximation.

When the perturber threshold leads to $\nu_p\ll \nu_r$ as is the case for Dy,
the Eq.~(\ref{eq:MQDTclo}) can be
approximated for energies where $\tan\pi\nu_p\sim -\tan\pi\delta_p$. For each of the
channels $p$ near their resonance, we approximate
In the neighborhood of the resonance
\begin{equation}
\tan\pi\nu_p = -\tan\pi\delta_p + (\epsilon - E_\alpha)\frac{d\tan\pi\nu}{d\epsilon}|_{\epsilon = E_\alpha}.
\end{equation}
to first order in $\epsilon - E_\alpha$. Trigonometric identities can be used to evaluate the
derivative:
\begin{equation}\label{eq:deriv}
\frac{d\tan\pi\nu}{d\epsilon}|_{\epsilon = E_\alpha} = \frac{\pi}{2}\frac{\nu_\alpha}{E_p-E_\alpha}(1+\tan^2\pi\nu_\alpha )
\end{equation}
where $E_p$ is the threshold energy for the excited core state and $E_\alpha$ is relative
to the ground core energy.

Using this approximation for the energy dependence, the exact Eq.~(\ref{eq:MQDTclo}) can
be approximated as Eq.~(\ref{eq:MQDTK}) with the identification
\begin{equation}
V_{i,\alpha} = K_{r_i,p_\alpha}/\sqrt{\frac{d\tan\pi\nu}{d\epsilon}|_{\epsilon = E_\alpha}}
\end{equation}

\section{Details of frame transformation}\label{sec:ftdetail}

If there is an angular momentum coupling
at small distances, $\beta$, that approximately commutes with the Hamiltonian
but coupling $b$ is more appropriate
when the electron is far from the core, then the
$K$-matrix in the $b$ representation is
\begin{equation}
    K_{b,b'}=\sum_\beta U_{b,\beta}K_\beta U^\dagger_{\beta ,b'}
\end{equation}
where $U_{b,\beta}=\langle b|\beta\rangle$.

For the $n$s series described in Sec.~\ref{sec:FT}, the transformation matrix
is determined by the recoupling of angular momentum. The $\beta$ type coupling gives
the three kets with the form $|(j_1(j_2,j_3)J_{23})J\rangle$ where $j_1$ is the total
angular momentum of the 4f electrons, $j_2$ and $j_3$ are the spins of the 6s and
$n$s electrons, and $J_{23}$ is the total spin coupling of the 6s and $n$s electrons:
$|(8(1/2,1/2)0)8\rangle$, $|(8(1/2,1/2)1)8\rangle$,
and $|(8(1/2,1/2)1)9\rangle$ since the orbital angular momentum from the 6s and
the $n$s is 0 for all couplings. The $b$ type coupling gives three kets with the form
$|((j_1,j_2)J_{12}j_3)J\rangle$ where the $j_1,j_2,j_3$ are as before and $J_{12}$ is the
total angular momentum of the core electrons (4f and 6s):
$|((8,1/2)15/2,1/2)8\rangle$, $|((8,1/2)17/2,1/2)8\rangle$, and $|((8,1/2)17/2,1/2)9\rangle$.
The $J=9$ only has one state in each coupling and therefore $U=1$. For $J=8$, the $U$
is a $2\times 2$ matrix given by the
projections $\langle b|\beta\rangle$ and is proportional to a 6$j$ coefficient:
\begin{eqnarray}
    U_{a,\alpha}&=&\langle ((j_1,j_2)J_{12}j_3)J|(j_1(j_2,j_3)J_{23})J\rangle\nonumber\\
    &=&(-1)^{j_1+j_2+j_3+J}[J_{12},J_{23}]
    \begin{Bmatrix}j_1 &j_2 &J_{12}\\ j_3& J&J_{23}\end{Bmatrix}
\end{eqnarray}
where $[J_{12},J_{23}]\equiv\sqrt{(2J_{12}+1)(2J_{23}+1)}$. If the $\beta = 1,2$ coupling are $J_{23}=0,1$
and the $b=1,2$ coupling are $J_{12}=17/2,15/2$, then
\begin{equation}
    U=\begin{pmatrix}
    \sqrt{\frac{9}{17}} & \sqrt{\frac{8}{17}} \\
    -\sqrt{\frac{8}{17}} & \sqrt{\frac{9}{17}} 
    \end{pmatrix}.
\end{equation}
Using this transformation matrix the four $K$-matrix elements for the $n$s series are
\begin{eqnarray}\label{eq:ftappr}
K^{(9)}_{17/2,17/2}&=&K_{^3S}\nonumber \\
K^{(8)}_{17/2,17/2}&=&\frac{9}{17}K_{^1S}+\frac{8}{17}K_{^3S}\nonumber\\
K^{(8)}_{15/2,15/2}&=&\frac{8}{17}K_{^1S}+\frac{9}{17}K_{^3S}\nonumber\\
K^{(8)}_{15/2,17/2}&=&-\frac{\sqrt{72}}{17}K_{^1S}+\frac{\sqrt{72}}{17}K_{^3S}
\end{eqnarray}

\bibliography{biblio.bib}

\end{document}